\documentclass[12pt,onecolumn]{article}%
\usepackage{amsmath}
\usepackage{graphicx}
\usepackage{amsfonts}
\usepackage{amssymb}%
\providecommand{\U}[1]{\protect\rule{.1in}{.1in}}
\begin{document}

\title{The Inner Life of the Kondo Ground State: \ An Answer to Kenneth Wilson's Question}
\author{Gerd Bergmann\\Department of Physics and Astronomy\\University of Southern California\\Los Angeles, California 90089-0484\\e-mail: bergmann@usc.edu}
\date{\today}
\maketitle

\begin{abstract}
The Kondo ground state has been investigated by numerical and exact methods,
but the physics behind these results remains veiled. Nobel prize winner
Wilson, who engineered the break through in his numerical renormalization
group theory, commented in his review article "the author has no simple
explanation ...for the crossover from weak to strong coupling". In this
article a graphical interpretation is given for the extraordinary properties
of the Kondo ground state. At the crossover all electron states in the low
energy range of $k_{B}T_{K}$ are synchronized. An internal orthogonality
catastrophe is averted.


\end{abstract}

\section{Introduction}

This year 2014 is the fiftieth anniversary of Kondo's \cite{K8} seminal paper
"Resistance minimum in Dilute Magnetic Alloys" and the fortieth anniversary of
Wilson's \cite{W18} renormalization paper about the Kondo effect. For 50 years
the Kondo effect has been investigated with the most sophisticated theoretical
methods \cite{A51}, \cite{F30}, \cite{K58}, \cite{K59}, \cite{N14}, \cite{N5},
\cite{N7}), \cite{G19}, \cite{B103}, \cite{W12}, \cite{A50}, \cite{S29},
\cite{B195}. Kondo \cite{K8} solved the puzzle of the low-temperature
resistance increase in dilute magnetic alloys \cite{H34} above the Kondo
temperature $T_{K}$. Wilson calculated the Kondo ground-state properties with
a numerical renormalization, known as NRG theory. He observed a crossover from
weak to strong coupling with increasing $n$ (number of renormalization steps).
In this article the FAIR solution of the Kondo ground-state \cite{B187} is
applied to reproduce and interpret Wilson's results (FAIR=Friedel artificially
inserted resonance).

\section{Wilson's Numerical Renormalization Theory}

The interaction between a magnetic impurity and the conduction electrons can
be described by an exchange interaction with the potential $-2J\left(
\mathbf{S\cdot s}\right)  \delta\left(  r\right)  $, $J<0$ where $\mathbf{S}$,
$\mathbf{s}$ are the spins of the impurity and the conduction electrons.
Wilson invented and applied a number of tricks to tackle the Kondo
ground-state. Using a band with a constant density of states and a band width
of $2W$ he divided all energies by $W,$ yielding a band range $\left(
-1:1\right)  $ with the Fermi level at $0$. Then Wilson made the (almost)
infinite number of s-electron states $\varphi_{k}^{\dagger}$ manageable by
dividing the band into energy cells. (I use the same symbol for a state, (for
example $\varphi_{k}$)$,$when addressing it by a creation operator
$\varphi_{k}^{\dagger}$, an annihilation operator $\widehat{\varphi}_{k}$ or
as a wave-function $\widetilde{\varphi}_{k}\left(  \mathbf{r}\right)  $).

In Fig.4 in the appendix Wilson's sub-division is shown. The ranges $\left(
-1:0\right)  $ and $\left(  0:1\right)  $ are split at $\pm1/2,\pm
1/4,\pm1/8,..\pm1/2^{\nu},..\pm1/2^{\infty}$. In the next step Wilson combined
all states $\varphi_{k}^{\dagger}$ within each cell $\mathfrak{C}_{\nu}$ into
a single state $c_{\nu}^{\dagger}$ (as the normalized sum of all states
$\varphi_{k}^{\dagger}$ within the cell). These states $c_{\nu}^{\dagger}$ I
will call Wilson states. They contain the full interaction of all electrons in
the cell with the impurity.

From these states $c_{\nu}^{\dagger}$ Wilson constructed a series of new
states $f_{\mu}^{\dagger}$. The state $f_{0}^{\dagger}$ is the normalized sum
of all the original band states $\varphi_{k}^{\dagger}$. It is concentrated at
the impurity, being a Wannier state of the s-band. The next state
$f_{1}^{\dagger}$ surrounds the inner state $f_{0}^{\dagger}$ and is itself
surrounded by $f_{2}^{\dagger}$, etc. All the $f_{\mu}^{\dagger}$ surround the
magnetic impurity like onion shells. Their width in real space increases each
time by a factor of two. Wilson chose the states $f_{\mu,\,\sigma}^{\dagger}$
in such a way that their Hamiltonian is that of a linear chain with next
nearest neighbor coupling. Only the states $f_{0,\sigma}^{\dagger}$ interact
with the impurity. He solved this Hamiltonian by renormalization, i.e. by
initially cutting off the chain at a small $n$ and solving the resulting
Hamiltonian $H_{n}$ by diagonalization. With the eigenstates of $H_{n}$ and
the states $f_{\left(  n+1\right)  \uparrow}^{\dagger},$ $f_{\left(
n+1\right)  \downarrow}^{\dagger}$ Wilson built the next Hamiltonian $H_{n+1}%
$. This NRG cycle is repeated. The number of basis states increases at each
NRG step by a factor of four (yielding $4^{n}$) but is generally limited to
the 1000 states with the lowest energies.

Wilson compared the resulting excitation spectrum for a finite exchange
interaction, for example $J=-0.055,$ with the spectrum for $J=0$ and
$J=-\infty$. For a small number $n$ of NRG steps the spectrum of $H_{n}$
resembled that of $J=0$. But after a critical number $n_{0}$ the spectrum
crossed over, resembling the strong coupling case $J=-\infty$. In addition,
Wilson observed that the effective number of band electrons changed from odd
to even at the transition.

With this work Wilson achieved a break through in the low-temperature
properties of the Kondo effect. From the flow diagram and the fixed-point
properties he obtained an effective Hamiltonian for low temperatures.
Evaluation of his numerical results lead Nozieres \cite{N14} to the
Fermi-liquid description of the Kondo ground-state.

Despite this great success, it appears that Wilson was not completely
satisfied with his achievement. In his review article about the Kondo
renormalization Wilson wrote (\cite{W18}, page 810): "Why the crossover from
weak to strong coupling takes place will not be explained. The author has no
simple physical explanation for it. It is the result of a complicated
numerical calculation".

The reason that Wilson had no simple interpretation of his results, i.e., that
the physics of the Kondo ground-state is so veiled, is due to the fact that
the wave function of the ground-state is so intangible. In NRG only a tiny
fraction of the ground-state Slater states can be maintained, which makes it
very difficult to uncover the hidden physics. Unfortunately the exact solution
using the Bethe-ansatz \cite{W12}, \cite{A50}, \cite{S29} does not help
because it is very difficult to extract the wave function from this ansatz.

\section{Magnetic and Kondo Ground-State in FAIR}

The author has developed in the past years a very compact solution for the
Kondo ground-state. It is known as the FAIR solution of the Kondo
ground-state. A short review is given in the festschrift to Jaques Friedel's
90's birthday \cite{B187} with extended references therein. Although it is not
an exact solution as the Bethe-ansatz, it describes the physics of the Kondo
ground-state very well, and it is well equipped to answer Wilson's implicit
questions behind the physics of the NRG cross-over.

Kondo and Wilson used a rigid magnetic impurity in their initial calculations.
However, the most common group of magnetic impurities are 3d-atoms dissolved
in a host. These impurities possess d-resonances. Friedel \cite{F57} and
Anderson \cite{A31} showed that a sufficiently large Coulomb exchange
interaction between opposite d-spins creates a magnetic moment in the
d-states. Anderson reduced the ten-fold degeneracy of the FA-impurity to a
two-fold degeneracy, making it de facto an impurity with $l=0$ and $s=1/2$ (it
is still called a d-impurity). This Anderson model is used in most theoretical
calculations of the Kondo effect of d-impurities. Schrieffer and Wolff
\cite{S31} showed that for sufficiently strong Coulomb interaction the
Anderson model yields the Kondo effect. Krishna-murthy, Wilkins, and Wilson
\cite{K58} performed NRG calculations for the FA-impurity and obtained an
equivalent crossover. Here I will discuss the Kondo ground-state of the
d-impurity because it demonstrates an interesting feed back of the singlet
state on the electronic structure of its magnetic components.

In the FAIR approach we use the same trick as Wilson to reduce the large
number of s-electron states. The positive and negative bands of s-electrons
are repeatedly sub-divided. But we stop the sub-division when a given number
$N=2n$ of energy cells $\mathfrak{C}_{\nu}\ $ is obtained, $n$ cells below and
$n$ cells above the Fermi level at energy zero. For each energy cell a Wilson
state is constructed. Then the smallest level spacing between the resulting
Wilson states is (next to the Fermi level) equal to $\delta=2^{-n+1}$ (in
units of $\varepsilon_{F}$ or $W$). The corresponding size of the host is
$R\thickapprox2^{n}\lambda_{F}/4$ where $\lambda_{F}$ is the Fermi wave
length. As in NRG the sample size \ doubles when $n$ is increased by one. Out
of the Wilson states two \emph{fair} states $a_{0\uparrow}^{\dagger}$ and
$b_{0\downarrow}^{\dagger}$ of spin-up and down are composed.

The easiest way to explain the logic behind the FAIR approach is to compare it
with a monarch whose subjects elect an ombudsman. This ombudsman does all the
negotiation with the king relieving all the other subjects from this duty. In
our case the $d_{\uparrow}^{\dagger}$-state is the king and the spin-up
s-states $c_{\nu\uparrow}^{\dagger}$ are the subjects. The latter elect the
\emph{fair} state $a_{0\uparrow}^{\dagger}$ as ombudsman who now exclusively
negotiates with the $d_{\uparrow}^{\dagger}$-state. This negotiation occurs in
form of \ s-d-hopping between $d_{\uparrow}^{\dagger}$ and $a_{0\uparrow
}^{\dagger}$, from which the remaining subjects $a_{i}^{\dagger}$ are
excluded. Their only function is to optimally elect and equip the ombudsman,
i.e. \emph{fair} state. (Of \ course, the remaining $\left(  N-1\right)  $
s-states $c_{\nu\uparrow}^{\dagger}$ have to be rebuilt so that they are
orthogonal to $a_{0\uparrow}^{\dagger}$, orthonormal to each other and
diagonal in the band Hamiltonian $H^{0}$. Because of the spin, there is a
second royal copy, consisting of $d_{\downarrow}^{\dagger}$ and
$b_{0\downarrow}^{\dagger}$. In the appendix the FAIR method is summarized for
a simple Friedel resonance.

This idea may appear too simple to work but actually it yields a much better
magnetic state for the Anderson model than the mean field theory \cite{B152}.
Equation (\ref{Psi_MS}) shows the structure of the magnetic state. For
sufficiently strong Coulomb interaction it assumes a magnetic moment, i.e.
$\left\vert B\right\vert ^{2}\neq\left\vert C\right\vert ^{2}$. \ For
$\left\vert B\right\vert ^{2}>>\left\vert A\right\vert ^{2},\left\vert
C\right\vert ^{2},\left\vert D\right\vert ^{2}$ the net d-spin is down. The
Coulomb repulsion affects only the term $Dd_{\uparrow}^{\dagger}d_{\downarrow
}^{\dagger}$ and the s-d-hopping is, for example, observed between the terms
$Ba_{0\uparrow}^{\dagger}d_{\downarrow}^{\dagger}$ and $Dd_{\uparrow}%
^{\dagger}d_{\downarrow}^{\dagger},$ $Aa_{0\uparrow}^{\dagger}b_{0\downarrow
}^{\dagger}$. The two half-filled FAIR bands $\left\vert \mathbf{0}%
_{a\uparrow}\right\rangle =a_{1\uparrow}^{\dagger}...a_{n\uparrow}^{\dagger
}\Omega$ and $\left\vert \mathbf{0}_{b\downarrow}\right\rangle =b_{1\downarrow
}^{\dagger}...b_{n\downarrow}^{\dagger}\Omega$ don't participate in any of the
interactions ($\Omega$ = vacuum state).%
\begin{equation}
\Psi_{MS\downarrow}=\left[  Aa_{0\uparrow}^{\dagger}b_{0\downarrow}^{\dagger
}+Ba_{0\uparrow}^{\dagger}d_{\downarrow}^{\dagger}+Cd_{\uparrow}^{\dagger
}b_{0\downarrow}^{\dagger}+Dd_{\uparrow}^{\dagger}d_{\downarrow}^{\dagger
}\right]  \left\vert \mathbf{0}_{a\uparrow}\right\rangle \left\vert
\mathbf{0}_{b\downarrow}\right\rangle \label{Psi_MS}%
\end{equation}

Fig.1 shows the electron structure of a magnetic d-impurity in the FAIR
description graphically. If one suppresses the spin-flip processes then one
obtains an enforced magnetic ground-state $\Psi_{MS\downarrow}$ with net
spin-down moment. The spin-up and -down FAIR bands are shown in the $\left\{
a_{i\uparrow}^{\dagger}\right\}  $- and $\left\{  b_{i\downarrow}^{\dagger
}\right\}  $-bases. The d-states are drawn to the left and right of the FAIR
bands. The circles within the FAIR bands represent the \emph{fair} states
$a_{0\uparrow}^{\dagger}$ and $b_{0\downarrow}^{\dagger},$ white is empty and
black is occupied. The figure shows the Slater state with the largest
amplitude. The double arrows indicate the transitions between the d- and the
\emph{fair} states via s-d-coupling. One obtains for the magnetic state a
total of four Slater states with the four possible occupations of d- and
\emph{fair} states as shown in equ. (\ref{Psi_MS}). The explicit form of the
magnetic solution is obtained by varying the composition of the two
\emph{fair} states $a_{0\uparrow}^{\dagger}$ and $b_{0\downarrow}^{\dagger}$
and minimizing the energy expectation value of the Anderson Hamiltonian. The
\emph{fair} states determine the remaining FAIR band states $a_{i\uparrow
}^{\dagger}$, $b_{i\downarrow}^{\dagger}$ and the coefficients $A,.,D$ uniquely.

Although the total spin of $\Psi_{MS\downarrow}$ in equ. (\ref{Psi_MS}) is
zero the d-impurity possesses a finite magnetic moment. The band electrons
which appear to compensate the moment are pushed to the surface of the host.%

\begin{align*}
&
%TCIMACRO{\FRAME{itbpF}{2.2756in}{2.0614in}{0in}{}{}{psi_{s}s_{s}%
%.eps}{\special{ language "Scientific Word";  type "GRAPHIC";
%width 2.2756in;  height 2.0614in;  depth 0in;  original-width 2.1461in;
%original-height 1.9419in;  cropleft "0";  croptop "1";  cropright "1";
%cropbottom "0";  filename 'Fig1.eps';file-properties "XNPEU";}}}%
%BeginExpansion
{\includegraphics[
height=2.0614in,
width=2.2756in
]%
{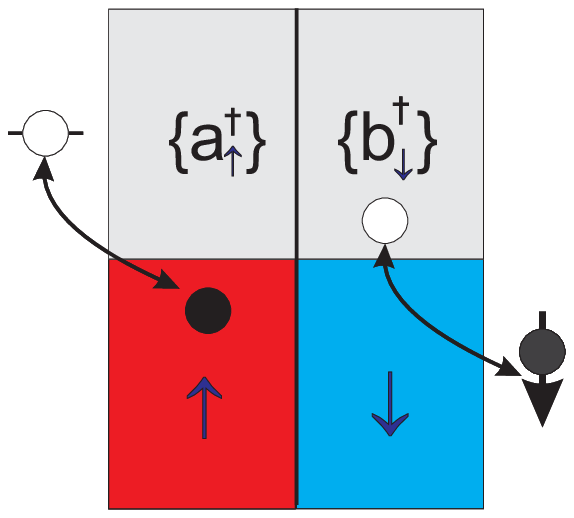}%
}%
%EndExpansion
\\
&
\begin{tabular}
[c]{l}%
Fig.1: The dominant Slater state for the magnetic state $\Psi_{MS\downarrow}$.
The spin of the\\
$\left\{  a_{i}^{\dagger}\right\}  $ FAIR band (red or dark) is anti-parallel
to the net spin of the magnetic state $\Psi_{MS\downarrow}$.
\end{tabular}
\end{align*}

If one reverses all spins in Fig.1 then one obtains an impurity $\Psi
_{MS\uparrow}$ with net spin up. (A modified version of) Both states together
will form the Kondo ground-state. But let us first consider the enforced
magnetic state $\Psi_{MS\downarrow}$ with a net d-spin down. Its half-filled
band states are $\left\vert \mathbf{0}_{a\uparrow}\right\rangle \left\vert
\mathbf{0}_{b\downarrow}\right\rangle $. Since the orbital wave functions
$a_{0}^{\dagger}$ and $b_{0}^{\dagger}$ of the \emph{fair} states are
different the corresponding bands $\left\{  a_{i}^{\dagger}\right\}  $ and
$\left\{  b_{j}^{\dagger}\right\}  $ are different too (the net spin of the
impurity breaks the up-down symmetry). Any transition between $\Psi
_{MS\downarrow}$ and $\Psi_{MS\uparrow}$ contains the multi-electron scalar
products (MESP)
\begin{equation}
\left\langle \mathbf{0}_{b\uparrow}\mathbf{0}_{a\downarrow}|\mathbf{0}%
_{a\uparrow}\mathbf{0}_{b\downarrow}\right\rangle =\left\langle \mathbf{0}%
_{b\downarrow}|\mathbf{0}_{a\downarrow}\right\rangle \left\langle
\mathbf{0}_{a\uparrow}|\mathbf{0}_{b\uparrow}\right\rangle =\left\vert
\left\langle \mathbf{0}_{a}|\mathbf{0}_{b}\right\rangle \right\vert ^{2}
\label{Dsp}%
\end{equation}

The MESP $\left\langle \mathbf{0}_{a}|\mathbf{0}_{b}\right\rangle $ is often
called the fidelity $F$. It can be calculated from the $\Psi_{MS\downarrow}$
alone if one takes only the orbital parts of $\left\vert \mathbf{0}%
_{a\uparrow}\right\rangle $ and $\left\vert \mathbf{0}_{b\downarrow
}\right\rangle $. The single electron states $a_{i\uparrow}^{\dagger}$ and
$b_{i\downarrow}^{\dagger}$ in Fig.1 experience the opposite polarization
potential. Therefore one expects that $\left\langle \mathbf{0}_{a}%
|\mathbf{0}_{b}\right\rangle $ in equ. (\ref{Dsp}) decreases with increasing
electron number or volume. It should show an orthogonality catastrophe.

\section{Internal Orthogonality Catastrophe}

We first check the fidelity $\left\langle \mathbf{0}_{a}|\mathbf{0}%
_{b}\right\rangle $ for the (enforced) magnetic state $\Psi_{MS\downarrow}$ as
a function of $n$ (where $n$ is half the number of Wilson states, $n=N/2$).
The smallest level spacing is $2^{-n+1}$ and the effective size of the host is
$2^{n}\lambda_{F}/4.$ In Fig.2 the logarithm of the fidelity $\ln\left(
F\right)  =\ln\left\langle \mathbf{0}_{a}|\mathbf{0}_{b}\right\rangle $ is
plotted as a function of $n$ for the magnetic state of a d-impurity (stars).
The parameters of the d-impurity are: d-state energy $E_{d}=-0.5,$ Coulomb
energy $U=1$ and s-d-hopping matrix element $\left\vert V_{sd}\right\vert
^{2}=0.03$. We find a linear dependence of $\ln\left(  F\right)  $ on $n$,
i.e., the fidelity $\left\langle \mathbf{0}_{a}|\mathbf{0}_{b}\right\rangle $
decreases exponentially with $n$. This causes an internal orthogonality
catastrophe (IOC), in analogy to Anderson's orthogonality catastrophe
\cite{A53}.

This IOC makes the transition matrix element between $\Psi_{MS\downarrow}$ and
$\Psi_{MS\uparrow}$ arbitrarily small for large sample volume and would
prevent any energy reduction in the singlet state. Therefore the IOC has to be
averted in the Kondo ground-state.

When spin-flip processes are permitted between $\Psi_{\downarrow}$ and
$\Psi_{\uparrow}$ then the system forms a singlet state. A new optimization
yields new compositions of the \emph{fair} states $a_{0}^{\dagger}$ and
$b_{0}^{\dagger}$ which yields new FAIR bands. The ground-state is the
(normalized) sum of the state in Fig.1 and its spin-inverted image. The
composition of $\Psi_{MS\downarrow}$ and $\Psi_{MS\uparrow}$ changes to a very
different form which I denote as $\Psi_{SS\downarrow}$ and $\Psi_{SS\uparrow}$
and the singlet ground-state is the normalized sum of $\Psi_{SS\downarrow}$
and $\Psi_{SS\uparrow}$. Now the fidelity shows a completely different
behavior (full circles in Fig.2). At about $n=15,$ the fidelity becomes
constant. The singlet state prevents the IOC. As we will see below this
transition into a constant fidelity at about $n=15$ is closely related to
Wilson's track change in in the NRG ladder.
\begin{align*}
&
%TCIMACRO{\FRAME{itbpF}{3.4338in}{2.8227in}{0in}{}{}{fig2.eps}%
%width 3.4338in;  height 2.8227in;  depth 0in;  original-width 3.8082in;
%original-height 3.1249in;  cropleft "0";  croptop "1";  cropright "1";
%cropbottom "0";  filename 'Fig2.eps';file-properties "XNPEU";}}}%
%BeginExpansion
{\includegraphics[
height=2.8227in,
width=3.4338in
]%
{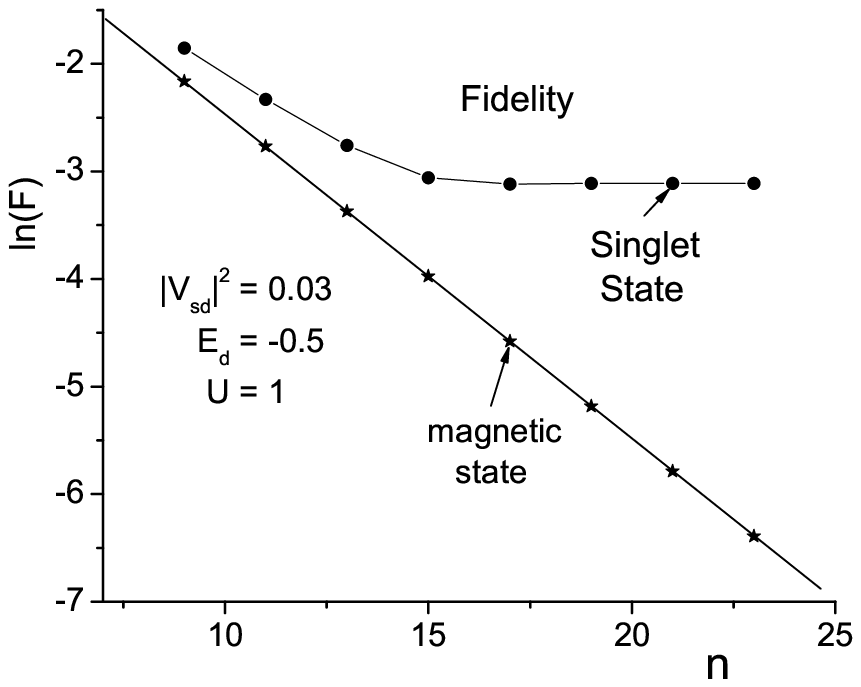}%
}%
%EndExpansion
\\
&
\begin{tabular}
[c]{l}%
Fig.2: The internal fidelity in the magnetic state (stars)\\
and the singlet state (full circles) as a function of $n.$\\
($2n$ is the number of Wilson states per spin. The radius\\
of the host is $2^{n}\lambda_{F}/4$).
\end{tabular}
\end{align*}

\section{Energy Shifts due to the Magnetic Impurity}

The formation of the singlet ground-state has a dramatic effect on the
electronic band structure. This becomes even more obvious when one
investigates the energy spectrum $E_{i}^{a}$ and $E_{i}^{b}$ of the two
FAIR-bands$.$ In the absence of the d-impurity the energy spectra for spin-up
and down are, of course, equal. We denote these initial energies as
$\varepsilon_{i}$. \{For the first $n-1$ states this energy depends
exponentially on $i$ and has the values $\varepsilon_{i}=-3/2\ast2^{-i}$,
while $\varepsilon_{n}=-2^{-n}$. Above the Fermi level one has the mirror
image of the negative energies. The total number of Wilson states for a given
$n$ is $N=2n.$ (The FAIR bands have one state less).

Now we can plot the relative energy shift $r_{i}=\left(  E_{i}-\varepsilon
_{i}\right)  /\left(  \varepsilon_{i+1}-\varepsilon_{i}\right)  $ for the two
FAIR bands, the $\left\{  a_{i}^{\dagger}\right\}  $- and the $\left\{
b_{i}^{\dagger}\right\}  $-band as a function of $i$. This is done in Fig.3.
The abscissa gives the number $i$ of the (energy ordered) Wilson states. The
second abscissa below is the corresponding energy scale.

We first discuss the energy shift in the magnetic state $\Psi_{MS\downarrow}$
(open triangles) (they are the same in $\Psi_{MS\uparrow}$). The increase of
$r_{i}$ from the left to the right of Fig.3 is due to the fact that the FAIR
bands have one state less than the band of Wilson states. The value of $r_{i}$
represents essentially the phase shift of the state $a_{i}^{\dagger}$ or
$b_{i}^{\dagger}$ in units of $\pi$. One recognizes that $r_{i},$ i.e. the
phase shifts are very different for the FAIR bands anti-parallel and parallel
to the net spin of the impurity. Close to the Fermi level the difference in
$r_{i}$ is almost equal to one, i.e. one level spacing.%
\begin{align*}
&
%TCIMACRO{\FRAME{itbpF}{3.9344in}{3.2594in}{0in}{}{}{fa_{r}in60_{e}%
%.eps}{\special{ language "Scientific Word";  type "GRAPHIC";
%width 3.9344in;  height 3.2594in;  depth 0in;  original-width 3.7293in;
%original-height 3.0843in;  cropleft "0";  croptop "1";  cropright "1";
%cropbottom "0";  filename 'Fig3.eps';file-properties "XNPEU";}}}%
%BeginExpansion
{\includegraphics[
height=3.2594in,
width=3.9344in
]%
{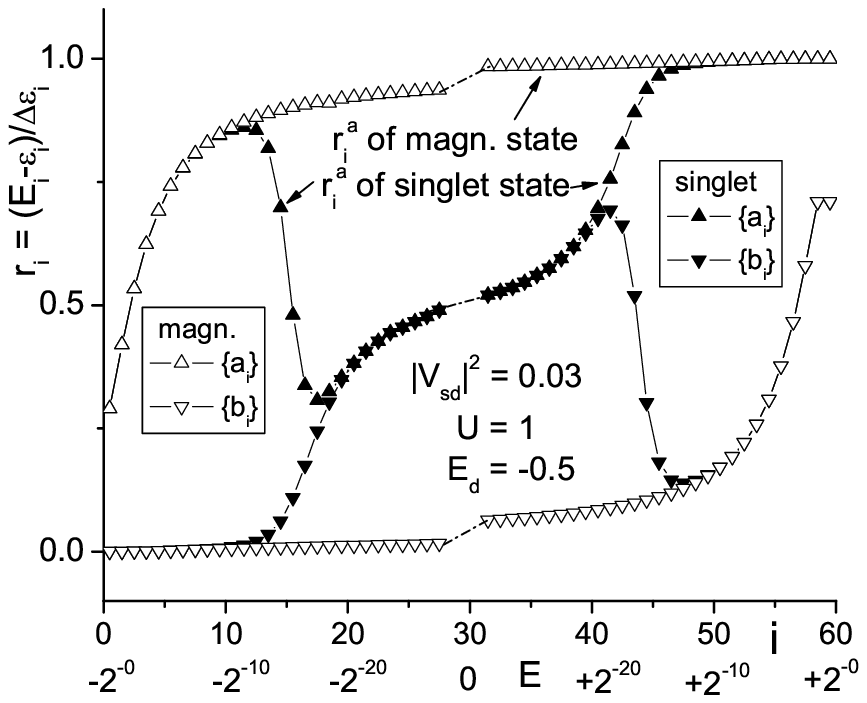}%
}%
%EndExpansion
\\
&
\begin{tabular}
[c]{l}%
Fig.3: The relative energy shifts $r_{i}$ of the FAIR bands $\left\{
a_{i}^{\dagger}\right\}  $ and $\left\{  b_{i}^{\dagger}\right\}  $\\
as a function of $i$ or energy (lower scale) for the magnetic state
$\Psi_{MS\uparrow}$\\
(open symbols) and the component $\Psi_{SS\uparrow}$ of the singlet state
(full symbols).
\end{tabular}
\end{align*}

In the singlet state the relative energy shift $r_{i}$, shown as full
triangles, presents a rather fascinating behavior. For $\left\vert
E\right\vert >2^{-10}$ the values of $r_{i}$ for the singlet and magnetic
states are quite close. However, if one approaches the low energy region,
$\left\vert E\right\vert <2^{-15}$, then the band energies $E_{i}^{a}$ and
$E_{i}^{b}$ move towards each other and become essentially identical. The
corresponding states $a_{i}^{\dagger}$ and $b_{i}^{\dagger}$ become
synchronized. As a consequence the internal orthogonality catastrophe is
averted in the Kondo ground-state.

The physical reason for the synchronization at low energy is the following. In
the Kondo ground-state one has a competition between polarization energy and
spin-flip energy. The spin-flip energy likes the two FAIR bands $\left\{
a_{i}^{\dagger}\right\}  $ and $\left\{  b_{i}^{\dagger}\right\}  $ to be
synchronized because its (transition) matrix element is proportional to
$\left\vert \left\langle \mathbf{0}_{a}|\mathbf{0}_{b}\right\rangle
\right\vert ^{2}.$ The polarization energy wants to shift the $\left\{
a_{i}^{\dagger}\right\}  $ and $\left\{  b_{i}^{\dagger}\right\}  $ bands in
opposite directions. At large (absolute) energies $\left\vert E\right\vert $
the polarization energy wins. Only for small energies of the order of
$k_{B}T_{K}$ (i.e. $n>n_{0}$) does the spin-flip gain a minor victory by
synchronizing the two bands in the very small energy range of the Kondo energy.

The synchronization of the two electron bases close to the Fermi level, i.e.
the suppression of the internal orthogonality catastrophe, is therefore a
characteristic property of the Kondo ground-state. In the process the two
\emph{fair }states $a_{0}^{\dagger}$ and $b_{0}^{\dagger}$ dramatically change
their composition. In the singlet state for $n>n_{0}$ they increase their
weight at very small energies.

Wilson's renormalization can be roughly visualized by means of the single
Fig.3. The figure corresponds to roughly $n=30$ NRG steps. If one wants to
visualize the situation after $10$ NRG steps one removes in Fig.3 the inner
section for $11\leq i\leq50$ and joins the remaining outer parts, then one
obtains, at least qualitatively, the relative energy shifts $r_{i}$ for
$n=10$. One easily recognizes that the crossover has not yet taken place. In
our case it occurs in the range $13<n<17$.

\section{The Physics of the Bound Electron in the Singlet State}

Wilson observed in his normalization sequence that \textbf{the spectrum
changed as if one electron was removed at the Fermi level }when the system had
crossed over from weak to strong coupling. The general interpretation is that
the impurity has bound one conduction electron and formed a singlet state,
removing this electron from the band.

In Fig.3 one recognizes that for the singlet state and $n=30$ the energies
$E_{i}^{a}$ and $E_{i}^{b}$ (for the same $i$ close to the Fermi level, i.e.
close to $n=30$) possess the same energy. An even number of band electrons
fills the two FAIR bands up to the same energy.

If one removes the inner forty states (as discussed above) then one obtains
roughly the energy shifts for $n=10$. Now the energy shifts $r_{i}$ for the
two FAIR bands in the singlet state differ roughly by one, $\Delta r_{i}%
=r_{i}^{a}-r_{i}^{b}$ $\thickapprox1$ or $\left(  E_{i}^{a}-E_{i}^{b}\right)
\thickapprox\left(  \varepsilon_{i+1}-\varepsilon_{i}\right)  $. Here the
energies $E_{i}^{a}$ lie about one level higher than $E_{i}^{b}$ (for the same
$i$). Now an odd number of electrons would fill the two FAIR bands up to the
same energy. This is exactly what Wilson observed. It is due to the
synchronization of the electron states within $k_{B}T_{K}$ of the Fermi level.
Of course, this is only possible when there are levels within $k_{B}T_{K}$ of
the Fermi level. For a spherical host this requires that the radius is larger
than the Kondo length.

\section{Summary}

In summary the synchronization of the FAIR band states close to the Fermi
level averts the internal orthogonality catastrophe between the states
$\Psi_{SS\uparrow}$ and $\Psi_{SS\downarrow}$. It arises because it permits
the system to lower its potential energy due to a tiny but finite spin-flip
energy between these two states. It is also this synchronization that appears
to remove an electron from the conduction band. At the same time the
composition of the $\emph{fair}$ states $a_{0}^{\dagger}$ and $b_{0}^{\dagger
}$ changes dramatically, which in turn changes the spectrum of the two FAIR
bands. This changes the charge distribution around each $\Psi_{SS\sigma}$
within a radius of the Kondo length, that is known as the Kondo cloud. It is
this low energy synchronization process which makes the Kondo effect such a
extraordinary phenomenon.

\textbf{This might be the physical interpretation Wilson was looking for.}

\newpage

\appendix{}

\section{Definition of Wilson states}

The ranges $\left(  -1:0\right)  $ and $\left(  0:1\right)  $ are split at
$\pm1/2,\pm1/4,\pm1/8,..\pm1/2^{\nu},..\pm1/2^{\infty}$. In the next step
Wilson combined all states $\varphi_{k}^{\dagger}$ within each cell
$\mathfrak{C}_{\nu}$ into a single state $c_{\nu}^{\dagger}$ (as the
normalized sum of all states $\varphi_{k}^{\dagger}$ within the cell). These
states $c_{\nu}^{\dagger}$ I will call Wilson states. They contain the full
interaction of all electrons in the cell with the impurity.%

\begin{align*}
&
%TCIMACRO{\FRAME{itbpF}{1.4388in}{2.9066in}{0in}{}{}{fig4.eps}%
%width 1.4388in;  height 2.9066in;  depth 0in;  original-width 2.1826in;
%original-height 4.4358in;  cropleft "0";  croptop "1";  cropright "1";
%cropbottom "0";  filename 'Fig4.eps';file-properties "XNPEU";}}}%
%BeginExpansion
{\includegraphics[
height=2.9066in,
width=1.4388in
]%
{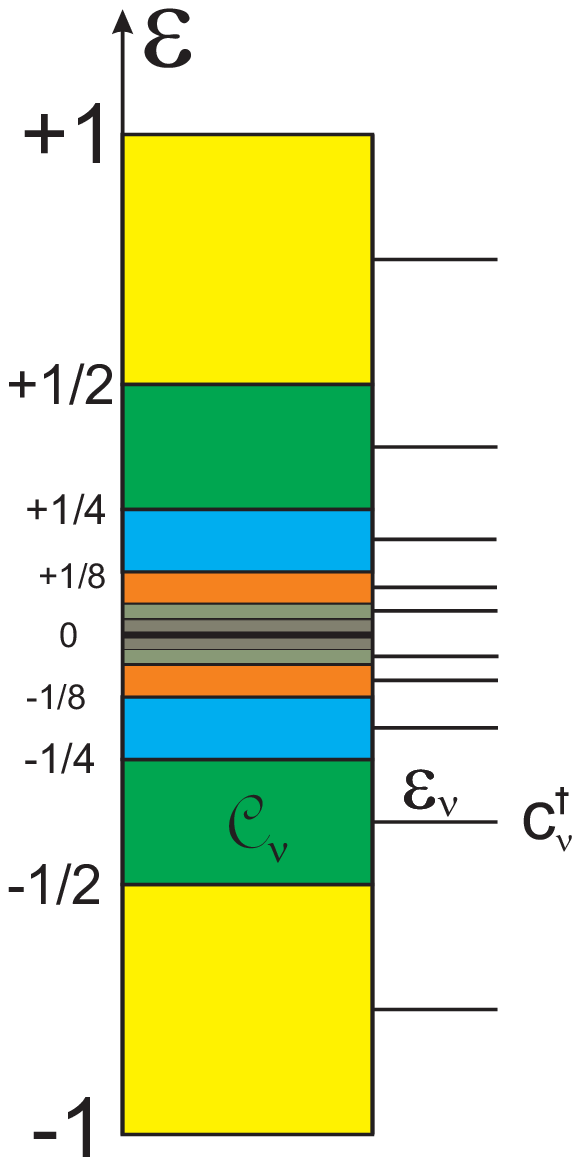}%
}%
%EndExpansion
\\
&
\begin{tabular}
[c]{l}%
Fig.4: The energy cells $\mathfrak{C}_{\nu}$, the\\
definition of the Wilson states $c_{\nu}^{\dagger}$\\
and their energies $\varepsilon_{\nu}$.
\end{tabular}
\end{align*}

\section{Summary of the FAIR method}

The FAIR ansatz can be best explained by the example of a spinless Friedel
resonance. In this case one has a (conduction) band of $N$ states $\left\{
c_{\nu}^{\dagger}\right\}  $ and a single (non-magnetic) d-impurity, the
so-called d-resonance. The d-state couples to every band state through the
s-d-matrix element $V_{sd}$. In the FAIR ansatz one constructs one \emph{fair
}state $a_{0}^{\dagger}$ out of the band states, i.e. as a normalized
superposition of band state wave functions. In the ground-state only this
\emph{fair} state interacts with the d-state. From the remaining $\left(
N-1\right)  $ band states a new FAIR band $\left\{  a_{i}^{\dagger}\right\}  $
is constructed. (First the $a_{i}^{\dagger}$ are made orthogonal to
$a_{0}^{\dagger}$ and orthonormalized. Then the band-Hamiltonian $\left(
H_{i,j}^{0}\right)  =\left(  \left\langle a_{i}^{\dagger}\Omega\left\vert
H^{0}\right\vert a_{j}^{\dagger}\Omega\right\rangle \right)  $ for $i,j>0$ is
diagonalized). The \emph{fair} state $a_{0}^{\dagger}$ is an artificial
Friedel resonance; it is coupled to each of the new band states $a_{i}%
^{\dagger}$ by a matrix element $V_{i}^{fr}$. The original band Hamiltonian
$H^{0}=%
%TCIMACRO{\dsum _{\nu}}%
%BeginExpansion
{\displaystyle\sum_{\nu}}
%EndExpansion
\varepsilon_{\nu}c_{\nu}^{\dagger}c_{\nu}$ is transformed into a Hamiltonian
with an (artificial) resonance state $a_{0}^{\dagger}$%
\[
H^{0}=%
%TCIMACRO{\dsum _{i=1}^{N-1}}%
%BeginExpansion
{\displaystyle\sum_{i=1}^{N-1}}
%EndExpansion
E_{i}a_{i}^{\dagger}a_{i}+E^{0}a_{0}^{\dagger}a_{0}+%
%TCIMACRO{\dsum _{i=1}^{N-1}}%
%BeginExpansion
{\displaystyle\sum_{i=1}^{N-1}}
%EndExpansion
V_{i}^{fr}\left(  a_{i}^{\dagger}a_{0}+a_{0}^{\dagger}a_{i}\right)
\]

In the ground-state the d-state couples only to the \emph{fair} state by the
matrix element $V_{0}^{sd}$ (which yields a $2\times2$ matrix) and the
ground-state for a half-filled band becomes%
\begin{equation}
\Psi_{FR}=\left(  Aa_{0}^{\dagger}+Bd^{\dagger}\right)
%TCIMACRO{\dprod \limits_{i=1}^{n-1}}%
%BeginExpansion
{\displaystyle\prod\limits_{i=1}^{n-1}}
%EndExpansion
a_{i}^{\dagger}\Omega\label{PsiFr}%
\end{equation}
Remarkably equ. (\ref{PsiFr}) represents the exact ground-state of the Friedel
resonance, and it has the advantage that it separates the states with zero and
one d-state.

For the Friedel resonance the composition of the \emph{fair} state
$a_{0}^{\dagger}$ is given by an exact formula. In other cases such as the
Kondo impurity one obtains the \emph{fair} states by variation (i.e.
minimizing the ground-state energy).

\section{The Kondo ground-state for $J=-\infty$, an exact FAIR solution}

Actually the wave function of the Kondo ground-state for $J=-\infty,$ which
Wilson and others used, is a very simple example of a FAIR solution. If for
example the impurity spin points down ($S_{\downarrow}$) then the term
$-2JS_{z}s_{z}\delta\left(  \mathbf{r}\right)  $ attracts all anti-parallel
spin-up states with a finite amplitude at $\mathbf{r=0}$ and builds out of all
band states a new state $\widetilde{a}_{0\uparrow}\left(  \mathbf{r}\right)  $
with the maximum amplitude at $\mathbf{r=0}$. It is given by
\[
\widetilde{a}_{0\uparrow}\left(  \mathbf{r}\right)  =\frac{1}{A}%
%TCIMACRO{\dsum _{\nu}}%
%BeginExpansion
{\displaystyle\sum_{\nu}}
%EndExpansion
\widetilde{c}_{\nu\uparrow}^{\ast}\left(  0\right)  \widetilde{c}_{\nu
\uparrow}\left(  \mathbf{r}\right)
\]
where $\widetilde{c}_{\nu\uparrow}^{\ast}\left(  \mathbf{0}\right)  $ is the
conjugate complex amplitude of the Wilson state $\widetilde{c}_{\nu\uparrow
}\left(  \mathbf{r}\right)  $ at $\mathbf{r=0}$ and $A$ is the renormalization factor.

If the band has $N$ states $\widetilde{c}_{\nu\uparrow}\left(  \mathbf{r}%
\right)  $ (or $c_{\nu\uparrow}^{\dagger}$) then the remaining $\left(
N-1\right)  $ states have to be rebuilt so that they are orthogonal to
$\widetilde{a}_{0\uparrow}\left(  \mathbf{r}\right)  $, orthonormal to each
other and diagonal in the band Hamiltonian $H^{0}$. We call this new band
$\left\{  a_{i\uparrow}^{\dagger}\right\}  $, $i>0$ a FAIR band. Actually the
impurity spin $S_{\downarrow}$ not only transforms the anti-parallel
conduction band but also the parallel one. In the latter any state with a
finite amplitude at $\mathbf{r=0}$ is forbidden. This means that all spin-down
band states have to be orthogonal to $\widetilde{a}_{0\downarrow}\left(
\mathbf{r}\right)  $. Since the orbital parts of $a_{0\uparrow}^{\dagger}$ and
$a_{0\downarrow}^{\dagger}$ are identical the corresponding band states
$a_{i\uparrow}^{\dagger}$ and $a_{i\downarrow}^{\dagger}$ possess the same
orbital wave functions.

If we ignore the spin-flip part of the exchange interaction for a moment then
we obtain the magnetic ground-state
\begin{equation}
\Psi_{MS,\downarrow}=S_{\downarrow}a_{0\uparrow}^{\dagger}\left\vert
\mathbf{0}_{a\uparrow}\right\rangle \left\vert \mathbf{0}_{a\downarrow
}\right\rangle \label{MS_J}%
\end{equation}
where $\left\vert \mathbf{0}_{a\uparrow}\right\rangle =%
%TCIMACRO{\dprod \limits_{i=1}^{n}}%
%BeginExpansion
{\displaystyle\prod\limits_{i=1}^{n}}
%EndExpansion
a_{i\uparrow}^{\dagger}\Omega$ represents the half-filled $\left\{
a_{i\uparrow}^{\dagger}\right\}  $-bands for spin up ($\Omega$ is the vacuum
state). The states $a_{0\uparrow}^{\dagger}$ and $a_{0\downarrow}^{\dagger}$
are artificial Friedel resonance states , denoted as \emph{fair} states. The
band states $\left\{  a_{i\uparrow}^{\dagger}\right\}  $ and $\left\{
a_{i\downarrow}^{\dagger}\right\}  ,$ ($i>0$) represent two new conduction
bands, the FAIR bands.

If one includes the spin-flip terms in the Hamiltonian then equ.
(\ref{J00_GS}) represents the ground-state of the Kondo impurity. It is a
simple version of a FAIR ground-state which is an exact solution for
$J=-\infty$.%
\begin{equation}
\Psi_{0}=\frac{1}{\sqrt{2}}\left(  S_{\downarrow}a_{0\uparrow}^{\dagger
}-S_{\uparrow}a_{0\downarrow}^{\dagger}\right)  \left\vert \mathbf{0}%
_{a\uparrow}\right\rangle \left\vert \mathbf{0}_{a\downarrow}\right\rangle
\label{J00_GS}%
\end{equation}
For $J=-\infty$ the anti-parallel and the parallel \emph{fair} states have the
same orbital wave function $\widetilde{a}_{0}\left(  \mathbf{r}\right)  $.
This is no longer the case for a finite value of $J$.

\section{The magnetic mean field solution}

It is \ worth noting that Anderson's mean field solution for the magnetic
state can be exactly expressed by a FAIR solution with the appropriate
\emph{fair} states. But it is not the optimal magnetic state. By optimizing
the two \emph{fair} states $a_{0}^{\dagger}$ and $b_{0}^{\dagger}$ one finds a
different magnetic solution which has a considerably lower ground-state energy
\cite{B152}. This FAIR solution requires twice the Coulomb exchange energy as
in mean-field theory to form a magnetic moment. The FAIR approach should be
included in spin-density functional theory calculations of the magnetic moment
of single impurities because its solution is superior to the presently applied
mean-field approximation.\newpage%

\[
\]

\end{document}